\documentclass[aps,12pt,a4paper]{revtex4}
\usepackage{epsfig}
\usepackage{graphicx}
\usepackage{amsmath,amssymb,color}
\usepackage[english]{babel}

\parskip=\medskipamount

\newcommand{\eq}[1]{(\ref{#1})}
\newcommand{\fig}[1]{Fig.~\ref{#1}}

\newcommand{\be}{\begin{equation}}
\newcommand{\ee}{\end{equation}}
\newcommand{\beq}{\begin{equation}}
\newcommand{\eeq}{\end{equation}}
\newcommand\disp{\displaystyle}

\newcommand{\la}{\left<}
\newcommand{\ra}{\right>}
\newcommand{\eps}{\varepsilon}

\begin{document}


\title{Equilibrium mean-field-like statistical models with KPZ scaling}

\author{Alexander Gorsky$^{1,2}$, Sergei Nechaev$^{3,4}$, and Alexander Valov$^{5}$}

\affiliation{$^1$Institute of Information Transmission Problems RAS, 127051 Moscow, Russia \\ $^2$ Moscow Institute of Physics and Technology, Dolgoprudny 141700, Russia \\ $^3$Interdisciplinary Scientific Center Poncelet (CNRS UMI 2615), 119002 Moscow, Russia \\ $^4$P.N. Lebedev Physical Institute RAS, 119991 Moscow, Russia \\ $^5$N.N. Semenov Institute of Chemical Physics RAS, 119991 Moscow, Russia}

\begin{abstract}

We have considered three different "one-body" statistical systems involving Brownian excursions, which possess for fluctuations Kardar-Parisi-Zhang scaling with the critical exponent $\nu=\frac{1}{3}$. In all models imposed external constraints push the underlying stochastic process to a large deviation regime. Specifically, we have considered fluctuations for: (i) Brownian excursions on non-uniform finite trees with linearly growing branching originating from the mean-field approximation of the Dumitriu-Edelman representation of matrix models, (ii) (1+1)D "magnetic" Dyck paths within the strip of finite width, (iii) inflated ideal polymer ring with fixed gyration radius. In the latter problem cutting off the long-ranged spatial fluctuations and leaving only the "typical" modes for stretched paths, we ensure the KPZ-like scaling for bond fluctuations. To the contrary, summing up all normal modes, we get the Gaussian behavior. In all considered models, KPZ fluctuations emerge in presence of two complementary conditions: (i) the trajectories are pushed to a large deviation region of a phase space, and (ii) the trajectories are leaning on an impenetrable boundary.

\end{abstract}

\maketitle

\section{Introduction}

Intensive investigation of extremal problems of correlated random variables in statistical mechanics has gradually lead mathematicians, and then, physicists, to the understanding that the Gaussian distribution is not as ubiquitous in nature, as it was supposed over the centuries, and shares its omnipresence with another universal law, known as the Tracy-Widom (TW) distribution. The "visiting card" of the TW law is the scaling exponent, $\nu$, of the second moment of the distribution, which is known as the Kardar-Parisi-Zhang (KPZ) exponent. For the first time this critical exponent was obtained in the seminal paper \cite{kpz} (see \cite{halpin} for review) for the nonequilibrium one-dimensional directed stochastic growth process. The theoretical analysis of growth was focused mainly on statistical properties of the enveloping surface growing in time, characterized by its hight, $h(x,t)$, where $x$ and $t$ are correspondingly the space and time coordinates. In 1+1 space-time dimensions, growing aggregate is characterized by the well-known scaling relation:
\be
\left({\rm Var}\,h(x,t)\right)^{1/2}=\frac{1}{N^{1/2}} \left(\sum_{x=1}^{N} \la h^2(x,t) \ra - \la
h(t) \ra^2 \right)^{1/2} = N^{1/2}g\left(\tau/N^{3/2}\right)
\label{eq01}
\ee
where the brackets denote averaging over different realizations of randomness during the growth, the variable $\tau=t/N$ is the averaged number of particles per one column of a growing aggregate, and the function $g(u)$ of the rescaled variable $u=\tau/N^{3/2}$ has the following asymptotic behavior:
\be
g(u)\sim \begin{cases} u^{1/3} & \mbox{for $u\ll 1$} \\
{\rm const} & \mbox{for $u\gg 1$} \end{cases}
\label{eq02}
\ee
Such a behavior is typical for many non-stationary processes, and the (1+1)-dimensional KPZ exponent, $\nu=\frac{1}{3}$, has been observed in a plenty of models of growth.

The breakthrough in understanding the ubiquity of KPZ statistics is connected with the works \cite{johansson,spohn} where it was realized that for flat initial conditions the distribution of a rescaled surface height, $\tau^{-1/3}(h(i,\tau)-2\tau)$, in a polynuclear growth converges as $\tau\to\infty$ to the Tracy-Widom (TW) distribution \cite{tw}, providing the statistics of edge states of random matrices belonging to the Gaussian Orthogonal Ensemble (GOE). Note that in the droplet geometry the statistics of growing surface instead corresponds to the edge states of the Gaussian Unitary Ensemble (GUE) \cite{spohn}. Simultaneously, it has been realized that the TW distribution describes the statistics of the ground state energy of an one-dimensional directed polymer in a random Gaussian potential and shortly later the Tracy-Widom distribution was re-derived using the replica formalism typical for disordered systems with the quenched uncorrelated disorder. It should be pointed out that the height fluctuations belonging to the KPZ universality class were measured also in experiments, both in planar \cite{miettinen, takeuchi1, takeuchi2} and in curved geometry in the electro-convection of nematic liquid crystals \cite{takeuchi1,takeuchi2} and a good quantitative agreement with TW distributions was reported.

One can ask a natural question: could we see some incarnations of a KPZ statistics in the equilibrium statistical mechanics besides the extremal events typical for the nonequlibrium growth? Namely, we perfectly know that the critical exponent, $\nu=\frac{1}{2}$, controls the fluctuations of the random walk, or, in the field-theoretic language, of the free particle quantum mechanics. So, we wonder whether one can find some simple one-body equilibrium statistical systems which possess fluctuations with the critical exponent $\mu = \frac{1}{3}$? The answer is positive and in next sections we provide examples of such models.

Despite the tremendous progress in understanding the mathematical background of the Tracy-Widom distribution and its relation to third-order phase transitions \cite{majumdar}, still, to our point of view, there is an essential lack in constructing clear and simple statistical models of the mean-field nature, which share the KPZ scaling and simultaneously shed light on the emergence of the third order phase transition. The bunch of works \cite{baruch, baruch2} have made a very significant contribution to the development of such models. These papers provided clear and transparent geometric optic approach for large deviations of statistics of Brownian trajectories pushed by external geometric constraints to an atypical region of the phase space.

Here we extend the line of reasoning of work \cite{baruch} and propose other exactly solvable one-body statistical models with the KPZ fluctuational behavior. Considering the paths counting problem on the "super" Cayley tree (the tree with the branching linearly depending on the tree generation), we discuss the KPZ-like scaling in the thermodynamic limit. We show that the model of path counting on supertrees can be regarded as the mean-field approximation of the Dumirtiu-Edelman view of the random matrix theory. The determinant representation of the Hermite polynomials is closely related to the characteristic polynomial for the transfer matrix on a supertree. When the "vertex degree velocity" (the branching increment between neighboring tree levels) is small, we can identify the corresponding model with the (1+1)D lattice random walk in the transverse constant magnetic field \cite{val}. The same problem can be formulated in the symmetric Riemann space with the non-constant radially-dependent curvature, which is the generalization of the space of constant negative curvature (the hyperbolic space). Finally, we connect the paths counting on the generalized supertrees with the statistics of one-dimensional Dyck paths with a fixed area below the path. This connection opens an interesting interpretation of paths ensembles on generalized supertrees in terms of construction of algebraic invariants of torus knots.

\section{Paths counting on finite supertrees}

The key object in eigenvalue statistics of random matrix ensembles is the joint eigenvalue distribution, $P_{\beta}(\lambda_1,...,\lambda_N)$,
\be
P_{\beta}(\lambda_1,...,\lambda_N) \propto \prod_{i\neq j}^N \left|\lambda_i-\lambda_j \right|^{\beta}\; e^{-c\sum\limits_{i=1}^N \lambda_i^2}
\label{eq03}
\ee
where $\beta$ depends on the matrix ensemble and typically takes the values $\beta = 1,2,4$ for Gaussian orthogonal, unitary and symplectic ensembles correspondingly.

In what follows, we pay attention to the one-point distribution functions, like, for example, the spectral density, $\rho(\lambda)$, which is defined as
\be
\rho(\lambda) = \frac{1}{N} \left<\sum_{i=1}^N\delta(\lambda - \lambda_i) \right> \propto
\int d\lambda_1 ... d\lambda_N \prod_{i\neq j}^N \left|\lambda_i-\lambda_j \right|^{\beta}\; e^{-c\sum\limits_{i=1}^N \lambda_i^2}\; \sum_{i=1}^N\delta(\lambda - \lambda_i)
\label{eq04}
\ee
At $N\gg 1$ in the bulk the spectral density is the celebrated Wigner semicircle, $\disp \rho(\lambda) = \frac{1}{\pi N}\sqrt{\lambda^2 - 4N}$, while near the spectral edge at $\lambda \simeq \lambda_{max}$, where $\lambda_{max} = 2\sqrt{N}$, it shares the Tracy-Widom distribution for the rescaled variable $\disp \xi = \frac{\lambda_{max} - 2 \sqrt{N}}{N^{-1/6}}$, such that
\be
\mathrm{Prob}(\xi<x) = e^{\disp -\int_x^{\infty} (s-x)\, q^2(s) ds}
\label{eq05}
\ee
where $x$ is $N$-independent, the function $q(s)$ satisfies the Painlev\'e II equation $q''(s) = 2q^3(s) + s q(s)$ with the asymptotic behavior $q(s)\big|_{s\gg1} \to \mathrm{Ai}(s)$, and $\disp \mathrm{Ai}(s)= \frac{1}{\pi} \int_{-\infty}^{\infty} \sin\left(\frac{u^3}{3}+s u\right)d u$. The result \eq{eq05} corresponds to the Gaussian Unitary Matrix (GUE) ensemble.

\subsection{Dumitriu-Edelman representation of matrix ensembles}

In the work \cite{edelman} I. Dumitriu and A. Edelman have shown that the spectral statistics of known matrix ensembles coincides with the spectral statistics of appropriately chosen ensembles of symmetric three-diagonal random matrices with independent matrix elements, uniformly distributed along the main diagonal, while non-uniformly distributed along two sub-diagonals. In particular, the joint distribution $P_{2}(\lambda_1,...,\lambda_N)$ of Gaussian Unitary Ensemble coincides with the spectral density of the ensemble of tri-diagonal symmetric matrices $M$, whose diagonal elements $x_{kk}$ ($k=1,...,K$) obey the normal distribution, $N(\mu,\sigma)$, while the sub-diagonal elements $x_{k,k+1}\equiv x_{k+k,i}$  ($k=1,...,K$) share the $\chi_{k}$-distribution. Remind that the normal and the $\chi$-distributions have the following probability densities for a random value, $x$:
\be
\left\{\begin{array}{rcll}
f(x|\mu,\sigma) & = & \disp \frac{1}{\sqrt{2\pi\sigma^2}}\, e^{-\frac{(x-\mu)^2}{2\sigma^2}} & \quad \mbox{for a normal, ${\cal N}(\mu,\sigma)$--distribution} \medskip \\
f(x|n) & = & \disp \frac{x^{k-1}e^{-\frac{x^2}{2}}}{2^{\frac{k}{2}-1} \Gamma\left(\frac{k}{2}\right)},\quad x\geq 0 & \quad \mbox{for a $\chi$--distribution}
\end{array}\right.
\label{eq06}
\ee
where $\Gamma(z)$ is the Gamma-function. Note that sub-diagonal elements $x_{k,k+1}$ have $k$-dependent probability densities $\chi_k$ and hence \emph{are not identically distributed}.

The symmetric matrix $M$ (with $x_{ij}=x_{ji}$) allows a straightforward interpretation as the transfer matrix of a path counting problem on a random symmetric tree \cite{val}. To make the statement more transparent, write $M$ together with the "shifted" matrix $M'$:
\be
M=\left(\begin{array}{cccccc}
x_{11} & x_{12} & 0 & 0 & 0 & \dots  \medskip \\
x_{21} & x_{22} & x_{23} & 0 & 0 & \medskip  \\
0 & x_{32} & x_{33} & x_{34} & 0 & \medskip \\
0 & 0 & x_{43} & x_{44} & x_{45} & \medskip \\
0 & 0 & 0 & x_{54} & x_{55} &  \medskip \\
\vdots & & & &  & \ddots
\end{array}
\right); \qquad
M'=\left(\begin{array}{cccccc}
x_{11} & 1 & 0 & 0 & 0 & \dots  \medskip \\
x^2_{21} & x_{22} & 1 & 0 & 0 & \medskip  \\
0 & x^2_{32} & x_{33} & 1 & 0 & \medskip \\
0 & 0 & x^2_{43} & x_{44} & 1 & \medskip \\
0 & 0 & 0 & x^2_{54} & x_{55} & \medskip \\
\vdots & & & & & \ddots
\end{array}
\right)
\label{eq07}
\ee
We can immediately see that for any distribution of matrix elements: $\det M = \det M'$. Below we deal with the matrix $\hat{M}'$ and consider it as a transfer matrix of a random tree constructed as follows:
\begin{itemize}
\item[(i)] All nodes at the generation $k$ of a tree carry one and the same $N(\mu,\sigma)$--distributed random weight;
\item[(ii)] The branching of all vertices in the generation $k$ of a tree is a $\chi_k$--distributed random variable.
\end{itemize}
For the path counting problem on such random tree, the condition (i) provides the normally distributed diagonal matrix elements, while (ii) ensures the $\chi$-distributed weights for passages between adjacent generations of the tree (from the generation $k$ to the generation $k+1$).

Define now a "mean tree", which is the result of the averaging over the ensemble of random trees constructed above. Instead of dealing with the spectral density of the ensemble of random matrices $M$, we study the eigenvalue distribution of the mean matrix $\la M \ra$, obtained by replacing each matrix element of $M$ by its mean value. The mean values of all diagonal elements are $0$ since the probability density, $f(x|\mu,\sigma)$, is symmetric at $\mu=0$, while the mean values (the expectations) of off-diagonal random elements are given by the following expression
\be
\mathbf{E}_{\chi_{(k)}}(x)= \frac{\sqrt{2}\,\Gamma\left(\frac{k+1}{2}\right)}{\Gamma\left(\frac{k}{2}\right)}
\label{eq08}
\ee
For $k\gg 1$ the expectation $\mathbf{E}_{\chi_{(k)}}(x)$ has the asymptotic behavior
\be
\mathbf{E}_{\chi_{(k)}}(x)\big|_{k\gg 1} = \sqrt{k}
\label{eq09}
\ee
Thus, the averaged matrices $\la M\ra$ and $\la M' \ra$ are:
\be
\la M\ra = \left(\begin{array}{cccccc}
0 & \sqrt{1} & 0 & 0 & 0 & \dots  \medskip \\
\sqrt{1} & 0 & \sqrt{2} & 0 & 0 & \medskip  \\
0 & \sqrt{2} & 0 & \sqrt{3} & 0 & \medskip \\
0 & 0 & \sqrt{3} & 0 & \sqrt{4} & \medskip \\
0 & 0 & 0 & \sqrt{4} & 0 & \\
\vdots & & & & & \ddots
\end{array}
\right)\quad \Rightarrow  \quad
\la M'\ra = \left(\begin{array}{cccccc}
0\;\; & 1\;\; & 0\;\; & 0\;\; & 0\;\; & \dots  \medskip \\
1\;\; & 0\;\; & 1\;\; & 0\;\; & 0\;\; & \medskip  \\
0\;\; & 2\;\; & 0\;\; & 1\;\; & 0\;\; & \medskip \\
0\;\; & 0\;\; & 3\;\; & 0\;\; & 1\;\; & \medskip \\
0\;\; & 0\;\; & 0\;\; & 4\;\; & 0\;\; & \medskip \\
\vdots & & & & & \ddots
\end{array}
\right)
\label{eq10}
\ee

\begin{figure}
\centering
\includegraphics[width=16cm]{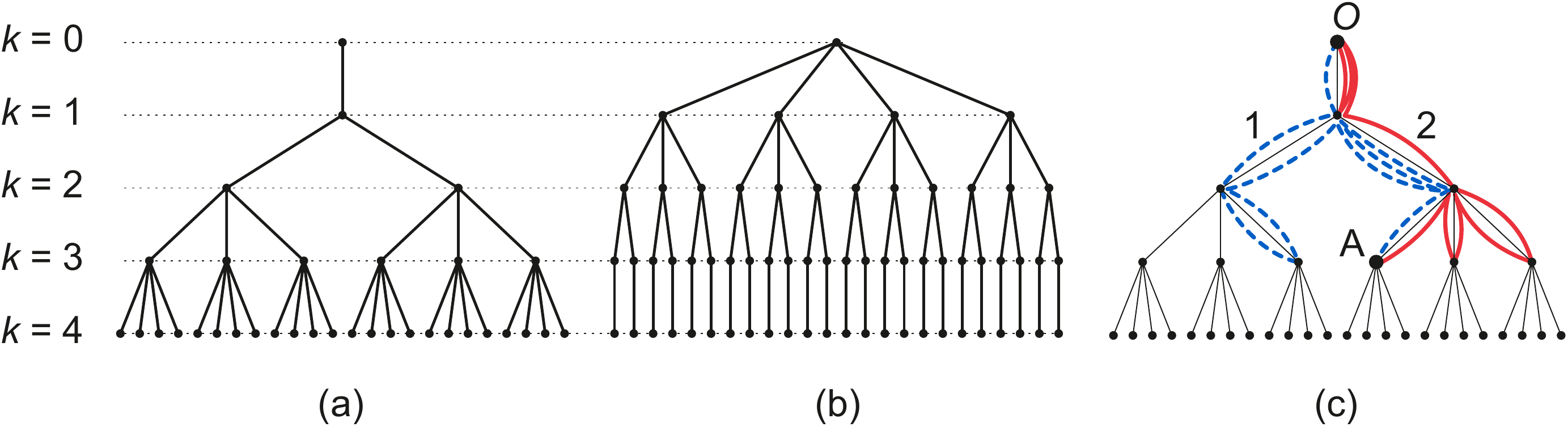}
\caption{Supertrees: (a) growing tree ${\cal T}^+$ with $p_0=1$ and $a=1$, (b) descending tree ${\cal T}^-$ with $p_0 = 4$ and $a = -1$, (c) Brownian bridge ("watermelon") configuration formed by two independent trajectories of $n=9$ steps each (solid and dashed). Both trajectories start from the root $0$ and join each other in the point $A$ on the level $k$ of an ascending tree.}
\label{fig01}
\end{figure}

The $K\times K$ matrix $\la M'\ra$ can be viewed as a transfer matrix of trajectories on symmetric ascending (${\cal T}^+$), or descending (${\cal T}^-$) "supertrees", of $K$ levels. The vertex degree, $p$, of these trees is not constant, but linearly depends on the current level, $k$ ($k=0,1,2,...,K-1$):
\be
p^+_k=\begin{cases} p^+_0, & \mbox{for $k=0$}  \medskip \\ 2+ak, & \mbox{for $k\ge 1$, $a > 0$} \end{cases} \qquad \mathrm{and} \qquad p^-_k=\begin{cases} p^-_0, & \mbox{for $k=0$}  \medskip \\ p^-_0-ak, & \mbox{for $k\ge 1$, $a > 0$} \end{cases}
\label{eq11}
\ee
for ${\cal T}^+$, and ${\cal T}^-$ supertrees, respectively as shown in \fig{fig01}a,b. The "branching velocity", $a$, is some integer-valued constant and $p^{\pm}_0$ are the branchings at the tree roots (at $k=0$). The matrix $\la M'\ra$ in \eq{eq10} considered upside-down corresponds to the ascending supertree ${\cal T}^+$ with $p_0^+=1$ and $a=1$, while regarding it bottoms-up we get the descending supertree ${\cal T}^-$ with $p_0^-=K$ and $a=-1$. The trees ${\cal T}^{\pm}$ are naturally to identify in the continuum limit with the symmetric Riemann spaces of non-constant negative curvature.

Before we proceed with random walks on supertrees, some important comment dealing with the paths counting on nonhomogeneous graphs should be made. Since the branching of the tree is not constant, we distinguish between the "path counting" (PC) problem and a more usual "random walk" (RW) statistics. The difference between PC and RW consists in different normalizations of the elementary step: for PC all steps enter in the partition function with the weight one, while for symmetric RW, the step probability depends on the current vertex degree, $p$: the probability to move along each graph bond equals $p^{-1}$. For graphs with a constant $p$ the PC partition function and the RW probability distribution differ only by the global normalization constant, and corresponding averages are indistinguishable. However, for inhomogeneous graphs, like supertrees ${\cal T}^{\pm}$, the distinction between PC and RW is crucial: in the path counting problem "entropic" localization of the paths may occur at vertices with large $p$, while it never happens for random walks. The distinction between PC and RW, and the entropic localization phenomenon were first reported for self-similar structures in \cite{17} and later were rediscovered for star graphs in \cite{ternovsky}. More recently this phenomenon was studied in \cite{burda} for regular lattices with defects, where authors introduced a notion of a "maximal entropy random walk" which is essentially identical to the path-counting problem. On a tree with one heavy root the localization phase transition in the path counting problem has been reported in \cite{heavy}.

\begin{figure}[ht]
\centering
\includegraphics[width=14cm]{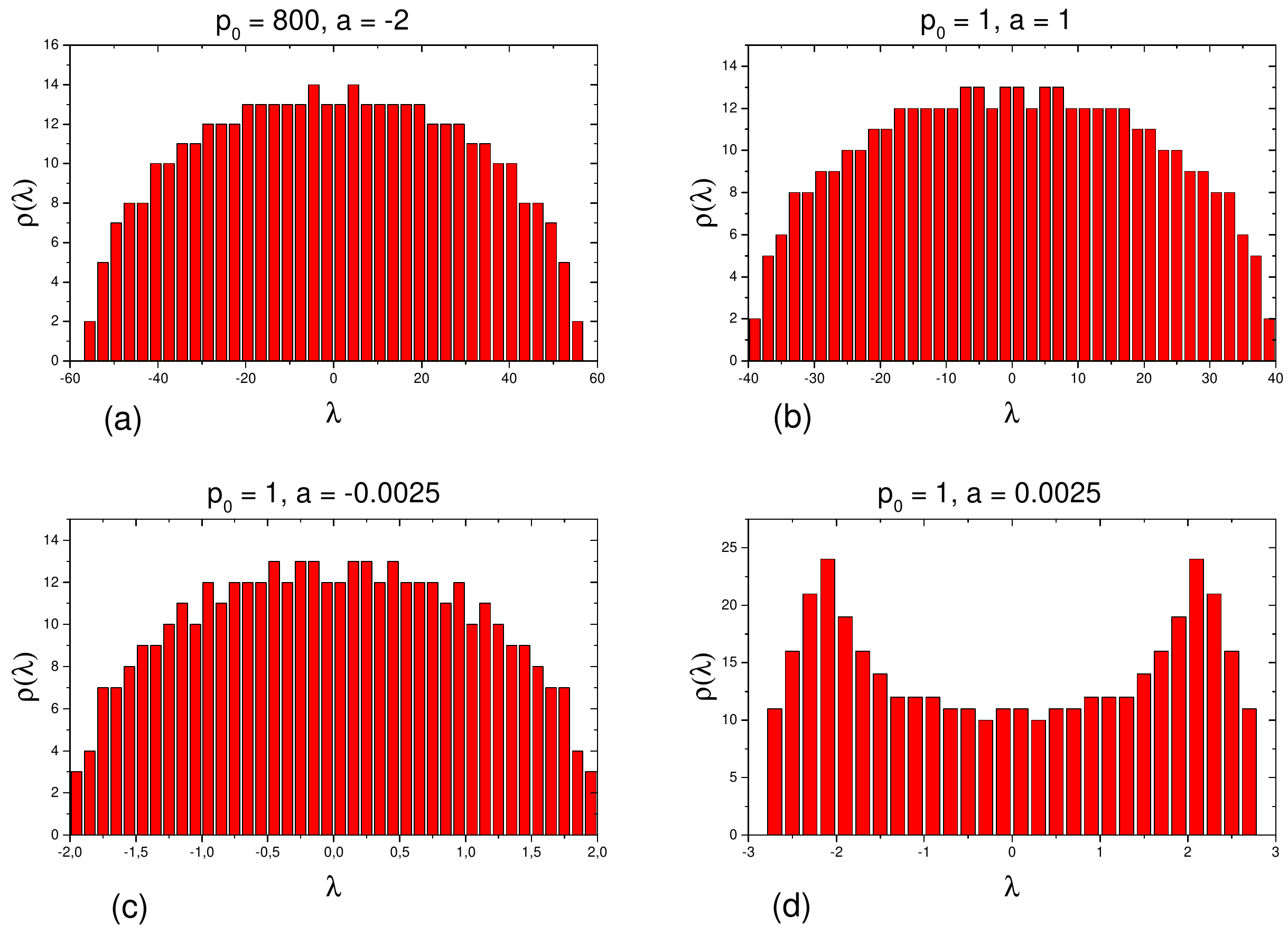}
\caption{Samples of spectral densities of transfer matrices for trees of $K=400$ generations and various cases: (a) $p_0=800, a=-2$; (b) $p_0=1, a=1$; (c) $p_0=1, a=-0.0025$; (d) $p_0=1, a=0.0025$.}
\label{fig02}
\end{figure}

Consider the following counting problem: given a regular finite tree, ${\cal T}$, compute the partition function, $Z_N(k)$, of all $N$-step trajectories starting at the tree root $(k=0)$ and ending at some tree level, $k$ ($k=0,...,K-1$). If ${\cal T}$ is the standard Cayley tree (or the Bethe lattice) with the constant branching, $p$, in each vertex at all tree levels, then this counting problem has been discussed infinitely many times in the literature in connection with various physical applications ranging from random walk statistics, polymer topology, localization phenomena, to questions dealing with the RG flows, holography and the black hole structure in the quantum field theory. In all mentioned cases, the uniform $p$--branching Cayley tree, is regarded as a discretization of the target space possessing the hyperbolic geometry -- the Riemann surface of the constant negative curvature. For a growing tree, the partition function, $Z_N(k)$, defined above, satisfies the recursion ($k=0,1,...,K-1$):
\be
\begin{cases}
Z_{N+1}(k)=(p_{k-1}-1)Z_N(k-1)+Z_N(k+1) & \mbox{for $2\le k \le K-1$} \medskip \\
Z_{N+1}(k)=Z_N(k+1), & \mbox{for $k=0$} \medskip \\
Z_{N+1}(k)=p_{k-1} Z_N(k-1)+Z_N(k+1) & \mbox{for $k=1$} \medskip \\
Z_{N+1}(k)=(p_{k-1}-1) Z_N(k-1), & \mbox{for $k=K-1$} \medskip \\
Z_{N=0} = \delta_{k,0}
\end{cases}
\label{eq12}
\ee
To rewrite \eq{eq12} in a matrix form, make a shift $k\to k+1$ and construct the $K$-dimensional vector $\mathbf{Z}_N=(Z_N(1),Z_N(2),... Z_N(K))^{\top}$. Then \eq{eq12} sets the evolution of $\mathbf{Z}_N$ in $N$:
\be
\mathbf{Z}_{N+1}=\hat{T}\mathbf{Z}_N; \qquad \hat{T}= \left(\begin{array}{cccccc}
0 & 1 & 0 & 0 &   \ldots & 0 \medskip \\ p_0 & 0 & 1 & 0 &  &  \medskip \\
0 & p_1-1 & 0 & 1 &  & \vdots \medskip \\ 0 & 0 & p_2-1 & 0 &  & \medskip \\
\vdots &  &  &  & \ddots &  \medskip \\ 0 &  & \dots & & p_{K-2}-1 & 0  \end{array}\right); \qquad \mathbf{Z}_{N=0}=\left(\begin{array}{c} 1 \medskip \\ 0 \medskip \\ 0 \medskip \\ 0 \medskip \\ \vdots \medskip \\ 0\end{array} \right)
\label{eq13}
\ee
Now we proceed in a standard way and diagonalize the matrix $\hat{T}$. The characteristic polynomials, $P_k(\lambda)=\det(\hat{T}-\lambda \hat{I})$, of the $k\times k$ matrix $\hat{T}$ satisfy the recursion
\be
\begin{cases}
P_k(\lambda)=-\lambda P_{k-1}(\lambda)-(p_{k-2}-1)P_{k-2}(\lambda), & \mbox{for $3\le k \le K$}
\medskip \\
P_1(\lambda)=-\lambda, \medskip \\
P_2(\lambda)=\lambda^2-p_0
\end{cases}
\label{eq14}
\ee
with $p_k$ given by \eq{eq11}. The spectral density, $\rho(\lambda)$, is constructed as follows. We solve the equation $P_K(\lambda)=0$ for a given $K$, get the set of eigenvalues $\{\lambda_1,...,\ lambda_K\}$ and construct the normalized histogram, which counts the degeneracies of each corresponding eigenvalue. The spectral densities for few different values of $p_0$ and $a$ are shown in \fig{fig02}. Specifically, we have plotted the $\rho(\lambda)$ for transfer matrices of size $K\times K$ for $K=400$ and the following sets of parameters: $p_0=800, a=-2$ for (a), $p_0=1, a=1$ for (b), $p_0=1, a=-0.0025$ for (c), and $p_0=1, a=0.0025$ for (d).

Now we discuss the analytic solution of \eq{eq12} for a growing tree ${\cal T}^+$ for a special choice $p_0=1$ and $a=1$, and analyze the corresponding asymptotics of $P_K$. The characteristic polynomials, $P_k(\lambda)$ of the transfer matrix $T$ satisfy the recursion \eq{eq14} with $p_0=1$, which coincide with the recursion for the so-called monic Hermite polynomials, ${\cal H}_k(\lambda)$, also known as the "probabilists' Hermite polynomials":
\be
P_k(\lambda) \equiv {\cal H}_k(\lambda)=(-1)^k e^{\frac{\lambda^2}{2}}{\frac{d^k} {d\lambda^k}}e^{-{\frac{\lambda^2}{2}}}; \qquad{\cal H}_k(\lambda)=2^{-k/2}H_k(\lambda / \sqrt{2)}
\label{eq15}
\ee
where $H_k(\lambda)$ are the standard Hermite polynomials. Hence the eigenvalues of the matrix $T$ of size $K\times K$ (see \eq{eq13}) are the roots of the monic Hermite polynomial, ${\cal H}_k(\lambda)$. In \cite{Kornyik} it has been shown that the normalized roots of the $K^{th}$ monic Hermite polynomial converge weakly at $K\gg 1$ to the Wigner semicircle,
\be
\rho(\lambda)=\frac{1}{2\pi K}\sqrt{4K-\lambda^2}
\label{eq16}
\ee
Thus, the spectral density $\rho(\lambda)$ has not any surprises at the spectral edge $|\lambda_{\max}|=2\sqrt{K}$.

Passing from the random matrix $M$ to the averaged one, $\la M \ra$, we have fully "washed out" the randomness and the following question arises: does the supertree carry any information about the KPZ scaling? Our claim is as follows: the signature of KPZ scaling is hidden is statistics of Brownian bridges (closed paths) on finite supertrees. Moreover, supposing the relation $N=cK$ between the number of steps $N$ of the Brownian bridge, and the size $K$ of the supertree, we can see the signature of the 3rd order phase transition (at $K\to \infty$) by varying $c$.

Specifically, we deal with the following problem: given a symmetric finite supertree $T^{\pm}$ of size $K$, we compute the conditional probability distribution, $W_N(k,n|K)$, of all $N$-step trajectories starting at the tree root $(k=0)$ and returning back after $N$ steps, to find an intermediate step $n$ at the distance $k$ along a tree. In particular we are interested in the distribution function $W_N\left(k,n=\frac{N}{2}|K\right)$, where $N$ is even. Computing
\be
\la \big(k-\la k \ra \big)^2\ra\sim K^{2\gamma}
\label{eq17}
\ee
in the thermodynamic limit $K\to \infty$ under the condition $N=cK$, we see different values of $\gamma$ below and above some $c^*$, namely:
\be
\gamma= \begin{cases} \frac{1}{2} & \mbox{for $c<c^*$} \medskip \\ \frac{1}{3} & \mbox{for $c>c^*$} \end{cases}
\label{eq18}
\ee

Since the Hermite polynomials are wave functions of the quantum oscillator we could wonder if the
KPZ scaling can be recognized in terms of this simplest quantum mechanical problem. To this aim
remind the representation of the oscillator in terms of the supertree and the matrix model \cite{krefl}. The wave function $\Psi_N(x)$ with $E_N= \hbar(N+\frac{1}{2})$ can be represented as the weighted sum of paths over supertree mentioned above
\beq
\Psi_N(x)\propto\sum_{{\rm paths}, N}e^{i\tilde{S}({\rm paths}|x)}
\label{path}
\eeq
where the sum runs over all paths on the supertree ending at $N$-th tree generation and the weight for each path is supposed to be $x$-dependent. The path integral \eq{path} can be considered as the dual path integral in the Hilbert space of the oscillator, where $x$ provides the weight oppositely to the conventional path representation, when the sum runs over the paths in the coordinate space and energy enters as the weight in the path integral. Two representations give one and the same answer for the partition function via the familiar identity for closed trajectories
\beq
Z(T)= \int_{{\rm paths}} dx(t)  e^{S(x(t))} = \mathrm{Tr}_{{\rm Hilbert\; space}}e^{-TH}
\eeq
where in the path integral the condition $x(t)=x(t+T)$ is imposed.

The related representation of the wave function in terms of the transfer-matrix is more
suitable for our purpose
\beq
\Psi_N(x)\propto \det(x-\hat{X})_{N\times N}
\eeq
where $\hat{X}= \sqrt{I}\cos(i\frac{d}{dI})$ is the operator of coordinate in the action-angle representation. At each step it provides a move along the radial direction on the phase space. This representation rhymes with the representation of the wave function in quantum mechanics via $N\times N$ Hermitian matrix $\beta$-ensemble with the Gaussian measure \cite{krefl}
\beq
\Psi_N(x) \propto \int \prod_i^{N} d\lambda_i (\lambda_i -\lambda_j)^{\beta} \det(M-x)^{\beta} e^{-\beta\, {\rm Tr}\, M^2}
\eeq
in the limit $\beta \to 0$, $\beta N=\mathrm{const}$, where this product defines the energy level $E=\hbar(\frac{1}{2} + \lim_{\beta\to 0} \beta N)$ The operator $\hat{X}$ amounts to the growth of the covered region in the phase space by adding the coherent states with $\hbar$ unit of area. The problem under consideration is formulated in energy space as follows. Consider the following
slightly unusual matrix element
\beq
W(N,K|n)\propto \langle 0|\hat{X}^N|K\rangle_n
\eeq
where the bra and ket vectors correspond to the wave functions in the action representation $\Psi(I)$. The operator $\hat{X}$ is the difference operator in this representation. The key point is that we introduce the cut-off in the action/energy space $I<n$ providing the restricted integration region in the matrix element. The question we are asking concerns a dependence of the matrix element with restricted integration area on the state $|K \rangle$ that is what is the typical variation around $\langle K \rangle$ for this matrix element as a function of cut-off n.

Note that in the conventional coordinate space the similar problem would sound as follows. Fix initial and final points of evolution $x_{{\rm in}},x_{{\rm out}}$ and consider the transition amplitude
\beq
\langle x_{{\rm out}}|\exp(i Ht)|x_{{\rm in}}\rangle
\eeq
where Hamiltonian is some differential or difference operator in the $x$-representation. Expand evolution operator, select $H^{N}$ term in the expansion and impose the restriction in coordinate space $x<X_0$. The function of interest would be the distributions of the final points of evolution as function of cut-off $X_0$ or equivalently $N$. Generally speaking we can consider any perturbation term $V$ in the Hamiltonian. In this case we would consider just possible final states
in the $N$-th order of perturbation theory in the theory with cut-off.

\subsection{Brownian bridges and KPZ statistics on finite supertrees}

The partition function, $Z_N(k)$, of $N$-step trajectories starting at the root point of a finite ascending tree ${\cal T}^+$ satisfies the recursion \eq{eq12}. To rewrite \eq{eq12} in a matrix form, make a shift $k\to k+1$ and construct the $K$-dimensional vector $\mathbf{Z}_N=(Z_N(1),Z_N(2),... Z_N(K))^{\top}$. Then \eq{eq13} sets the evolution of $\mathbf{Z}_N$ in $N$:
\be
\mathbf{Z}_{N+1}=T_{K\times K}\mathbf{Z}_N; \qquad T_{K\times K}= \left(\begin{array}{cccccc}
0 & 1 & 0 & 0 &  \ldots & 0 \smallskip \\ 1 & 0 & 1 & 0 &  &  \smallskip \\
0 & 2 & 0 & 1 &  & \smallskip \\ 0 & 0 & 3 & 0 &  & \smallskip \\
\vdots &  &  &  & \ddots &  \smallskip \\ 0 &  & \dots & & K-1 & 0  \end{array}\right); \qquad \mathbf{Z}_{N=0}=\left(\begin{array}{c} 1 \smallskip \\ 0 \smallskip \\ 0 \smallskip \\ 0 \smallskip \\ \vdots \smallskip \\ 0\end{array} \right)
\label{eq19}
\ee
The partition function $Z_N(k)$ of $N$-step paths ending at the level $k$ of the supertree, can be straightforwardly expressed via the matrix element of the $N^{\rm th}$ power of the transfer matrix, $Z_N(k)=\left(T^N_{K\times K}\right)_{1,k}$. Let $N$ and $K$ be even and $k$ is odd ($k=2m+1$,  $m\in \left[0, \frac{P}{2}-1\right]$). At $N>K\gg 1$ we have
\be
Z_N(k)\approx \lambda_{max}^N \frac{{\cal H}_{k-1}(\lambda_{max})}{\disp \sum_{j=1}^{K}\frac{{\cal H}^2_{j-1}(\lambda_{max})}{(j-1)!}}
\label{eq20}
\ee
where ${\cal H}_k(\lambda)$ is the so-called monic Hermite polynomial (see \eq{eq15}). At $N\gg 1$ the dominant contribution to the partition function is given by the largest eigenvalue $\lambda_{max}$ of the transfer-matrix $T_{K\times K}$. Thus, at $N\gg 1$ the probability to find the path's end at the level $k$ of a supertree can be estimated as
\be
P_N(k) \approx \frac{H_{k-1}(\lambda_{max})}{\sum\limits_{i=0}^{K/2-1}H_{2i}(\lambda_{max})}
\label{eq21}
\ee

Now we can compute the conditional probability, $Q_N(k,n|K)$, to find the $n$'th step of the path at the level $k$ under the condition that at the $N$'th step the path returns to the tree root and the whole tree has $K$ generations. The corresponding "watermelon" (i.e. conditional Brownian bridge) configuration consists of two trajectories, 1 and 2, of lengths $n$ and $N-n$, both starting at the root point $0$ -- see \fig{fig01}c, and meeting each other in the point $A$ located at the distance $k$ along a finite tree of $K$ levels. The function $Q_N(k,n|K)$ reads
\be
Q_N(k,n|K)=C_0\,\frac{P_n(k)P_{N-n}(k)}{k!P_N(0)} =C_0\,\frac{{\cal H}^2_{k-1}(\lambda_{max})}{k!}
\label{eq22}
\ee
where $C_0=\left(\sum_k \frac{{\cal H}^2_{k-1}(\lambda_{max})}{k!}\right)^{-1}$. Without the loss of generality and for simplicity we take $n=\frac{N}{2}$, i.e. we consider the distribution of the middle point of the Brownian bridge and in \eq{eq20} it is supposed that $n = N-n > K$.

The behavior of monic Hermite polynomials, ${\cal H}_k(\lambda)$, in the vicinity of the spectral edge has been analyzed in \cite{Dominici}. At $\lambda\approx 2\sqrt{K}$, the polynomials ${\cal H}_k(\lambda)$ share the following asymptotics
\be
{\cal H}_K(\lambda)\approx \sqrt{2\pi}\, 2^{-K/2} \exp\left(\frac{K\ln(2K)}{2}- \frac{3K}{2}+ \lambda\sqrt{K}\right) K^{1/6}\mathrm{Ai}\left(\frac{\lambda-2\sqrt{K}}{K^{-1/6}}\right)
\label{eq23}
\ee
where $\disp {\rm Ai}(z)=\frac{1}{\pi} \int_{0}^{\infty} \cos(\xi^3/3+\xi z)\, d\xi$ is the Airy function. Let $a_1>a_2>a_3>...$ be zeros of the Airy function ($a_i<0$ for all $i$). At $K\gg 1$ the maximal eigenvalue, $\lambda_{max}$, of the transfer matrix \eq{eq19} has the following leading behavior
\be
\lambda_{max}=2\sqrt{K}+a_1\, K^{-1/6}
\label{eq24}
\ee
where $a_1\approx -2.3381$ is the first zero of Airy functon.

\begin{figure}[ht]
\centering
\includegraphics[width=15cm]{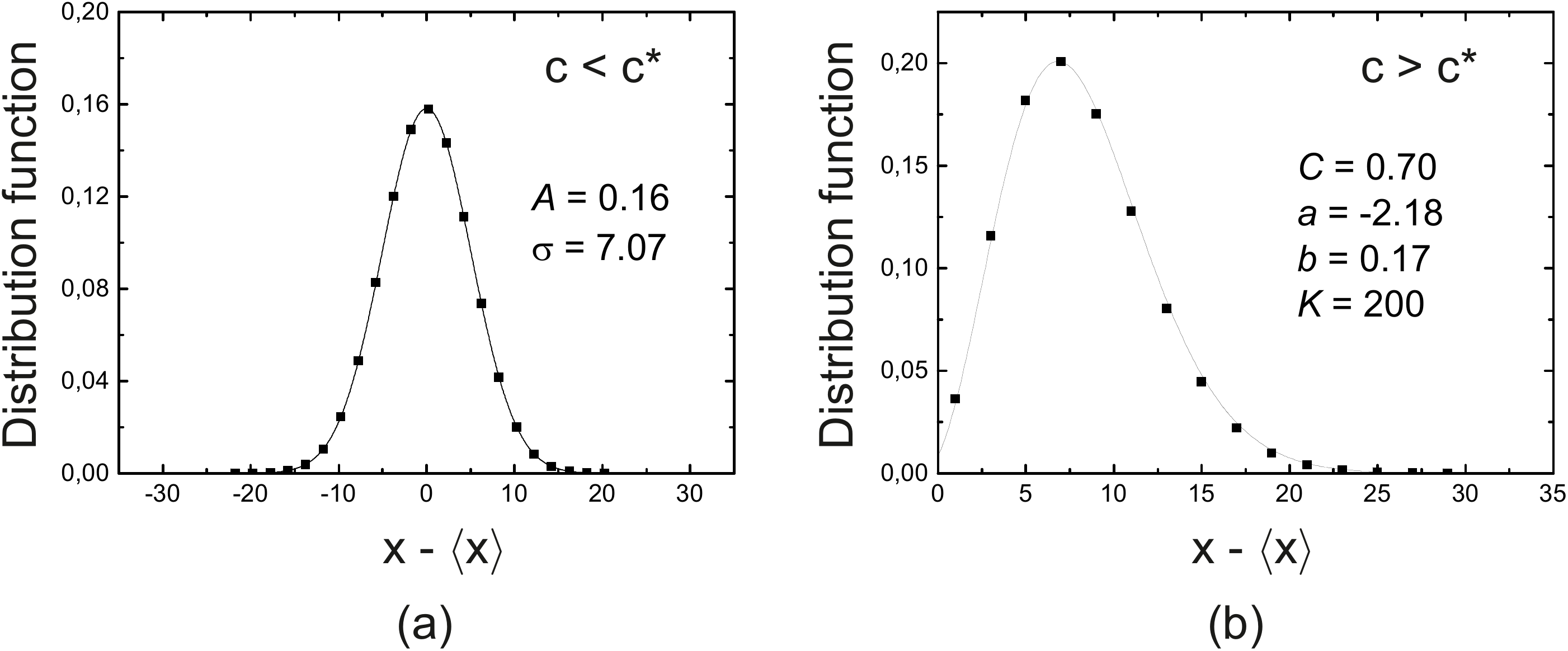}
\caption{Distribution functions of the middle points of Brownian bridges: (a)  Comparison of the shifted distribution $Q_N(k,n|K)$ at $c<c^*$ against Gaussian distribution $A \exp \left(-\frac{(x-\langle x \rangle)^2}{\sigma^2}\right)$ with $A=0.16, \sigma = 7.07$; (b) Shifted distribution $Q_N(k,n|K)$ at $c>c^*$ against the Airy-type distribution $C Ai^2(a+b(K-x))$ with $C=0.70, a=-2.18, b=0.17, K=200$, and $N=8K$ (the parameters $a,b$ for given $K$ are defined in \eq{eq25a}).}
\label{fig03}
\end{figure}

Let us suppose that the the number of steps in the Brownian bridge, $N$, is a linear function of the maximal tree size, $K$, i.e. $N=c K$. Making the substitution $k=K-j$ at $\frac{j}{K}\ll 1$, we get for $c>c^*$ the following limiting behavior:
\be
Q_N(K-j,n=N/2|K)\Big|_{N=cK}=\frac{{\cal H}^2_{K-j-1}(\lambda_{max})}{(K-j)!}\approx C^{-1} \mathrm{Ai}^2\left(a+b j\right)
\label{eq25}
\ee
where $C=\sum_i \mathrm{Ai}^2(a+b j)$ and the constants $a$ and $b$ are defined as follows:
\be
a=\frac{1}{K^{1/3}} - a_1\left(1-\frac{1}{6K}\right), \qquad
b=\frac{1}{K^{1/3}}- \frac{a_1}{6K}
\label{eq25a}
\ee
For $c>c^*$, where $c^*\approx 4$, the square root of the variance $\sqrt{\mathrm{Var}\, k} = \sqrt{\mathbb{E}(k-\mathbb{E}k)^2}$ of the conditional distribution function $Q_{N=cK}$ has the following asymptotic behavior at large $K$:
\be
\sqrt{\mathrm{Var}\, k} \sim K^\gamma, \quad \gamma=\frac{1}{3}
\label{eq26}
\ee
For $c<c^*$ the conditional distribution function $Q_{N=cK}$ is Gaussian and the corresponding variance at large $K$ reads
\be
\sqrt{\mathrm{Var}\, k} \sim K^\gamma, \quad \gamma=\frac{1}{2}
\label{eq27}
\ee
The distribution functions of the middle poits of Brownian bridges at $c<c^*$ and at $c>c^*$, obtained on the basis of transfer matrix computations are plotted in \fig{fig03} together with their fits.

\begin{figure}[ht]
\centering
\includegraphics[width=15.5cm]{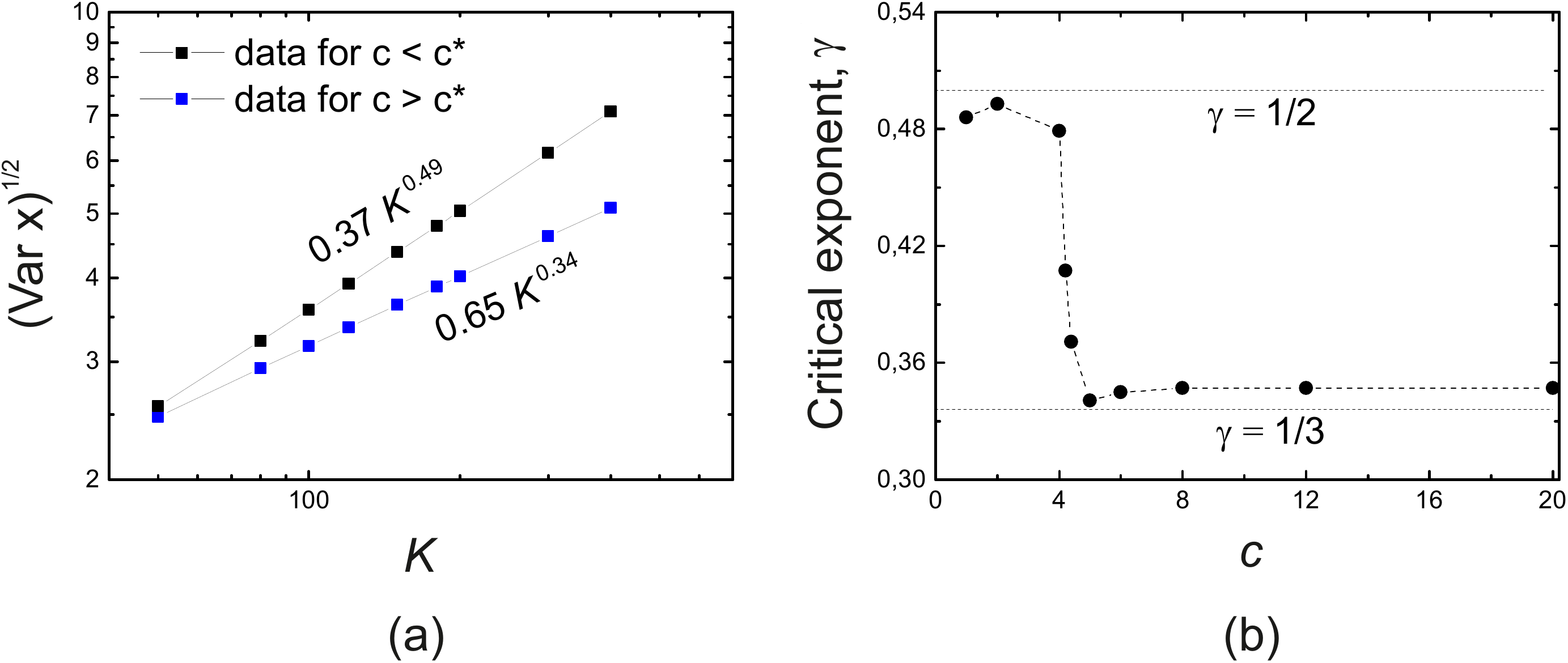}
\caption{(a) Fluctuations of the Brownian bridge middle point in the "compressed" regime at $c>c^*$; (b) Critical exponents for Brownian bridge fluctuations on finite supertrees at various $c=N/K$. The phase transition at $c=c^*\approx 4$ is clearly seen.}
 \label{fig04}
\end{figure}

The system experiences at $c^*$ the phase transition which is manifested in different fluctuation regimes below and above $c^*$. The corresponding phase diagram is depicted in \fig{fig04}.

In \fig{fig04} the critical exponent $\gamma$ is plotted against $c$, which quantitatively measures how strong the Brownian bridge is interacting with the supertree boundary: at $c<2$ the Brownan bridge does not feel the boundary at all, while at $c>2$ the tree "compresses" the Brownian bridge. Note that the transition occurs at $c=c^*\approx 4$, which means that to reach the Airy-type distribution with KPZ-type fluctuations, the Brownian bridge of length $N$ should be sufficiently squeezed on a finite supertree of $K$ generations ($N>c^* K$).

\section{Magnetic Dyck paths in a strip as random walks on supertrees with small "branching velocity"}

In the previous section we have discussed the tree with $a=1$ branching velocity. Here we consider the limit of small branching velocity $|a|\ll 1$. In \cite{val} we have argued that the random walk on such trees is closely related to the area-weighted Dyck paths, where the effective magnetic field on the lattice is related to the branching velocity.

\begin{figure}[ht]
\centering
\includegraphics[width=12cm]{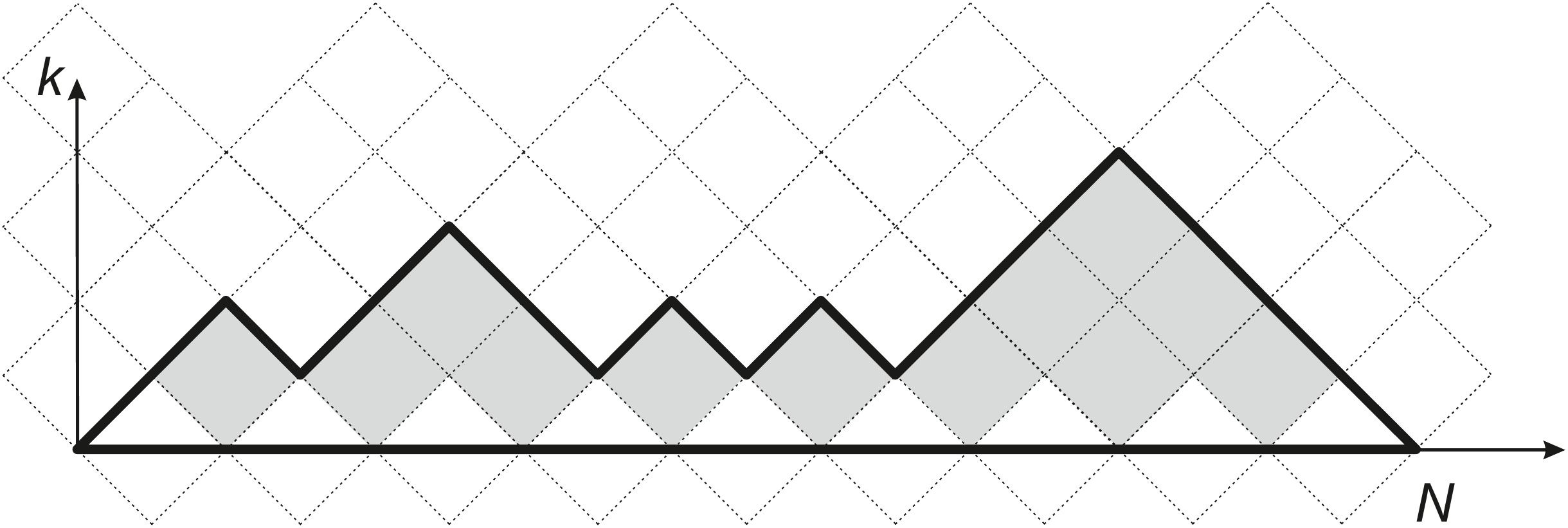}
\caption{The $N$-step Dyck path on a halfline $k\ge 0$ with fixed area below the path measured in full plaquettes.}
\label{fig05}
\end{figure}

Consider a $N\times N$ square lattice and enumerate all $N$-step trajectories (Dyck paths) starting at $(0,0)$, ending at $(N,N)$ and staying above the diagonal of the square (the path can touch the diagonal, but cannot cross it). Let $A$ be the area between the path and the diagonal of the square, counted in full plaquettes. For convenience, turn the lattice by $\pi/4$ and consider the partition function of all directed $N$-step paths on a half-line, $k\ge 0$, with fixed area, $A$, being the sum of all full gray plaquettes as shown in \fig{fig05}. Our key object is the area-weighted canonical partition function, $W_N(q)$, defined as follows
\be
Z_N(q) = \sum_{\rm Dyck\; paths} q^A
\label{eq28}
\ee
where the summation runs over the ensemble of $N$-step Dyck paths enclosing the area $A$, and $q$
is the fugacity of $A$. Writing $q=e^H$, we identify $H$ with a "magnetic field" conjugated to the area $A$.

Let us introduce the partition function $Z_N(k,q)$ where $k$ is the height of the path at step $N$. The function $Z_N(k,q)$ satisfies the recursion
\be
\begin{cases}
Z_{N+1}(k)=Z_N(k-1)+q^k Z_N(k+1) & \mbox{for $0< k<\infty$} \medskip \\
Z_{N+1}(k)=Z_N(k+1), & \mbox{for $k=0$} \medskip \\
Z_{N=0}(k) = \delta_{k,0}
\end{cases}
\label{eq29}
\ee
Note that the value $Z_N(0,q)$ defines the partition function of the "Brownian excursion", since at the very last step the trajectory returns to the starting point. Evaluating powers of the matrix $U(q)$, we can straightforwardly check that the values of $Z_N(0,q)$ are given by the Carlitz-Riordan $q$-Catalan numbers \cite{carlitz}:
\be
Z_N(0,q) = \begin{cases} C_{N/2}(q) & \mbox{for $N=2m$, where $m=1,2,3...$} \medskip \\ 0 & \mbox{for $N=2m+1$, where $m=0,1,2,...$}
\end{cases}
\label{eq30}
\ee
Recall that the numbers $C_N(q)$ satisfy the recursion
\be
C_N(q) = \sum_{k=0}^{N-1}q^k C_k(q) C_{N-k-1}(q)
\label{eq31}
\ee
which is the $q$-extension of the standard recursion for Catalan numbers. The generating function
$\disp F(s,q)=\sum_{N=0}^{\infty}s^N C_N(q)$ obeys the functional relation
\be
F(s,q) = 1 + s F(s,q) F(sq,q)
\label{eq32}
\ee
It is known that the solution of \eq{eq32} can be written as a continuous fraction expansion,
\be
F(s,q) =\frac{1}{\disp 1-\frac{s}{\disp 1-\frac{s q}{\disp 1- \frac{s q^2}{1-...}}}}
=\frac{A_q(s)}{A_q(s/q)}
\label{eq33}
\ee
where $A_q(s)$ is the $q$-Airy function,
\be
A_q(s)=\sum_{n=0}^{\infty}\frac{q^{n^2}(-s)^n}{(q;q)_n}; \quad (t;q)_n=\prod_{k=0}^{n-1} (1-t q^k)
\label{eq34}
\ee
In the works \cite{prellberg0,rich1,rich2} it has been shown that in the double scaling limit $q\to 1^-$ and $s\to \frac{1}{4}^{-}$ the function $F(s,q)$ has the following asymptotic form (compare to \eq{eq33})
\be
F(z,q) \sim {\cal F}_{\rm reg}+(1-q)^{1/3} \frac{d}{dz}\ln {\rm Ai}(4z); \quad
z=\frac{\frac{1}{4}-s}{(1-q)^{2/3}},
\label{eq35}
\ee
where $F_{\rm reg}$ is the regular part at $\big(q\to 1^-,\, s\to \frac{1}{4}^{-}\big)$. The function ${\cal F}(s,1)$ is the generating function for the non-deformed Catalan numbers:
\be
F(s,q=1)=\frac{1-\sqrt{1-4s}}{2s}
\label{eq36}
\ee
The generating function $F(s,1)$ is defined for $0<s<\frac{1}{4}$, and at the point $s=\frac{1}{4}$ the first derivative of $F(s,1)$ experiences a singularity which is interpreted as the critical behavior. The limit $q\to 1^-$, $s\to\frac{1}{4}^-$ can be read also from the asymptotic expression for $F(s,q)$. To define the double scaling behavior and derive the Airy-type asymptotic, the simultaneous scaling in $s$ and $q$ is required.

To make connection of area-weighted Brownian excursions to the path counting problem on descending tree, consider the expansion of \eq{eq29} at $q\to 1$. Namely we set $q=1-\eps$, where $|\eps|\ll 1$ and expand \eq{eq29} up to the first term in $\eps$. We arrive at the following system of equations \cite{val}
\be
\begin{cases}
Z_{N+1}(k,q) = \big(1-\eps(k-1)\big) Z_N(k-1,q) + Z_N(k+1,q) & \mbox{for $1\le k\le K-1$} \medskip \\ Z_{N=0}(k,q)=\delta_{k,0}
\end{cases}
\label{eq37}
\ee
Comparing equations \eq{eq37} and \eq{eq12} we can note that they are equivalent upon the identification $a= -\eps$ (where $0<\eps\ll 1$) and $p_0=1$. Hence, \eq{eq37} provides the explicit expression for the path counting on a weakly descending tree with a small non-integer branching velocity, $a$, by expanding the solution to \eq{eq29} at $q=1-\eps$ up to the first leading term in $\eps$ and an identification $\eps$ with $a$.

Let us introduce the partition function $Z_N(k,q)$ in a strip where $k$ is the height of the path at step $N$ and $k$ bounded by the size of a strip $n$. The function $Z_N(k,q)$ satisfies the recursion
\be
\begin{cases}
Z_{N+1}(k)=Z_N(k-1)+q^k Z_N(k+1) & \mbox{for $0< k\leq n$} \medskip \\
Z_{N+1}(k)=Z_N(k+1), & \mbox{for $k=0$} \medskip \\
Z_{N=0}(k) = \delta_{k,0}
\end{cases}
\label{eq29b}
\ee
and consider the generating function
\be
Z(s,k)=\sum\limits_{N=0}^{\infty}Z_N(k)s^k
\label{eq38}
\ee
For $Z(s,k)$ the recursion relations \eq{eq29b} can be rewritten as follows
\be
\mathbf{Z}_{\delta}=\hat{T}_{(n+1)\times (n+1)}\mathbf{Z}_s;
\label{eq39}
\ee
where
\be
\hat{T}_{(n+1)\times (n+1)}= \left(\begin{array}{cccccc}
s^{-1} & -1 & 0 & 0 & \ldots & 0 \medskip \\ -1 & s^{-1} & -q & 0 & \ldots & 0 \medskip \\ 0 & -1 & s^{-1} & -q^2 & \medskip \\ \vdots & \vdots & \ddots & \ddots & \ddots & \medskip  \\ 0 & 0 &  & -1 & s^{-1} & -q^{n-1} \medskip \\ 0 & 0 &  & 0 & -1 & s^{-1} \end{array}\right);  \mathbf{Z}_{\delta}=\left(\begin{array}{c} s^{-1} \medskip \\ 0 \medskip \\ 0 \medskip \\  \vdots \medskip \\ 0\end{array} \right);
 \mathbf{Z}_{s}=\left(\begin{array}{c} Z_s(0) \medskip \\ Z_s(1) \medskip \\ Z_s(2) \medskip \\  \vdots \medskip \\ Z_s(k)\end{array} \right)
\label{eq40}
\ee
The generating function of the number of area-weighted Dyck paths returning to $k=0$ in the finite-width case has been obtained in \cite{prellberg}:
\be
Z_s(k=0)=\frac{\sum\limits_{m=0}^{\left[n/2\right] }(-s^2)^m q^{m^2}\bigl[\!\begin{smallmatrix} n-m \\ m \end{smallmatrix}\!\bigr]_q}{\sum\limits_{m=0}^{\left[(n+1)/2\right] }(-s^2)^m q^{m(m-1)}\bigl[\!\begin{smallmatrix} n+1-m \\ m \end{smallmatrix}\!\bigr]_q},
\label{eq41}
\ee
where
\be
\bigl[\!\begin{smallmatrix} n \\ m \end{smallmatrix}\!\bigr]_q=\frac{(q;q)_n}{(q;q)_m (q;q)_{n-m}}
\label{eq42}
\ee
We provide generalization of \eq{eq41} and present generic expression of the generating function of magnetic Dyck paths which end at the level $k$
\be
Z_s(k)=\frac{s^k\sum\limits_{m=0}^{\left[( n-k )/2\right] }(-s^2)^m q^{m(m+k)}\bigl[\!\begin{smallmatrix}   n-k-m  \\ m \end{smallmatrix}\!\bigr]_q}{\sum\limits_{m=0}^{\left[(n+1)/2\right] }(-s^2)^m q^{m(m-1)}\bigl[\!\begin{smallmatrix} n+1-m \\ m \end{smallmatrix}\!\bigr]_q}; \qquad 0\leq k\leq n
\label{eq43}
\ee

In the space $(N,q)$ the recursion relations \eq{eq29b} rewritten in the matrix form with a shift $n+1\rightarrow n$, read
\be
\mathbf{Z}_{N+1}=T_{n\times n}\mathbf{Z}_{N};
\label{eq44}
\ee
where
\be
T_{n\times n}= \left(\begin{array}{cccccc}
0 & 1 & 0 & 0 & \ldots & 0 \medskip \\ 1 & 0 & q & 0 & \ldots & 0 \medskip \\ 0 & 1 & 0 & q^2 & \medskip \\ \vdots & \vdots & \ddots & \ddots & \ddots & \medskip  \\ 0 & 0 &  & 1 & 0 & q^{n-2} \medskip \\ 0 & 0 &  & 0 & 1 & 0 \end{array}\right);  \mathbf{Z}_{0}=\left(\begin{array}{c} 1 \medskip \\ 0 \medskip \\ 0 \medskip \\  \vdots \medskip \\ 0\end{array} \right);
\label{eq45}
\ee
Following the receipt of \cite{quassem}, we can use the following representation of $\mathbf{Z}_{N}$:
\be
Z_{N}(k)=T_{1,k}^N=\sum\limits_{m=1}^{n} x_m^N \frac{p_{k-1}(x_m)\gamma_n(k)}{N_n(x_m)} \qquad k=\overline{1,n},
\label{eq46}
\ee
where
\be
\gamma_n(k)=\prod_{i=k}^{n-1} q^{i-1}; \qquad N_n(x)=\sum\limits_{i=1}^n \gamma_n(i) p^2_{i-1}(x);
\label{eq47}
\ee
and
\be
p_n(x)=\det (T_{n\times n} - x I)=(-x)^n\sum\limits_{m=0}^{\left[n/2\right] }(-x^{-2})^m q^{m(m-1)}\bigl[\!\begin{smallmatrix}   n-m  \\ m \end{smallmatrix}\!\bigr]_q,
\label{eq48}
\ee
and $\{ x_k\}_{k=\overline{1,n}}$ are the eigenvalues of the matrix $T_{n\times n}$ (или $p_n(x_k)=0$). For $N\gg n$ the following approximation is valid:
\be
Z_N(k) \approx 2 x^N_{max} \frac{p_{k-1}(x_{max})\gamma_n(k)}{N_n(x_{max})};
\label{eq50}
\ee
and the probability to find the end of the trajectory at the level $k$ is independent on $N$.

\begin{figure}[ht]
\centering
\includegraphics[width=10cm]{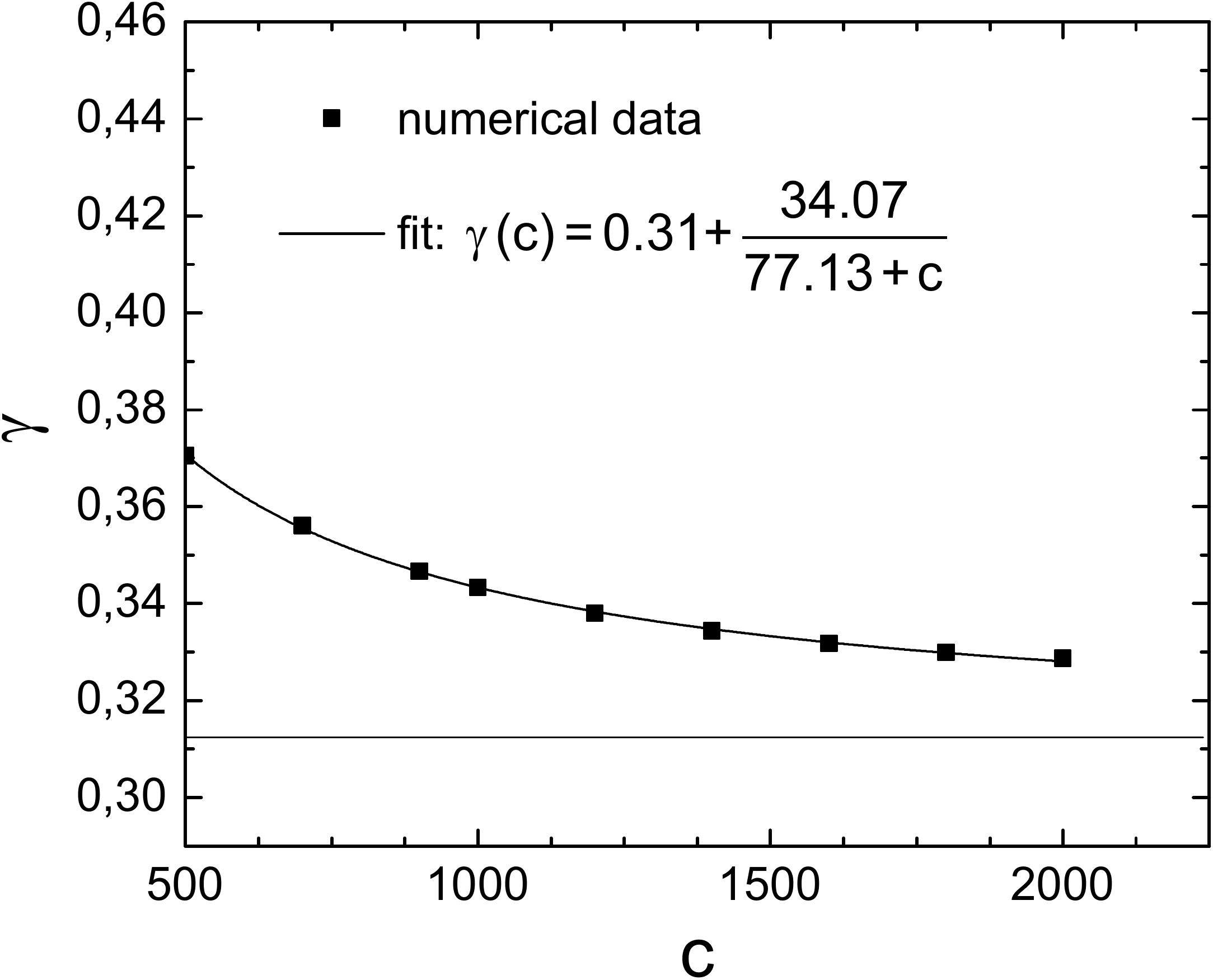}
\caption{Critical exponent $\gamma(c)$ in the fluctuation of the middle point of inflated $2N$-step Dyck path with $q=1+N^{-2/3}$ in a strip of width $n$ at strong stretching, $N = cn$.}
\label{fig06}
\end{figure}

The conditional probability to find the middle point of the $2N$ step Brownian bridge at the level $k$:
\be
Q(N,k)=\frac{Z^2_N(k)}{\sum\limits_{k=0}^n Z^2_N(k)}\approx\frac{p^2_{k-1}(x_{max})\gamma^2_n(k)}{\sum\limits_{k=1}^n p^2_{k-1}(x_{max})\gamma^2_n(k)}
\label{eq50a}
\ee
Expression \eq{eq50a} permits to compute height fluctuations, $\sqrt{\langle (k-\langle k \rangle)^2\rangle}$ of the middle point located at the height $k$, as a function of a strip size, $n$, in slightly inflated Dyck paths, when the inflation is controlled by the length-depended fugacity, $q=1+N^{-\beta}$ ($\beta>0$). For some $\beta$ we expect the universal scaling,
\be
\mathrm{Var}\, k(n) = \la (k-\langle k \rangle)^2\ra = \disp \sum_{k=0}^n k^2 Q(N,k)-\left(\sum_{k=0}^n k Q(N,k)\right)^2 \sim n^{2\gamma (c)}
\label{eq50b}
\ee
where the condition $N = cn$ is supposed, i.e. the number of steps, $N$, of Dyck path scales linearly with the strip width, $n$. The condition $N = cn$ is chosen since it rhymes with the relation $N = cK$ in the dependence between the number of steps, $N$, of the Brownian excursion and the maximal number of supertree generations, $K$, imposed in Section IIA (see \eq{eq17}-\eq{eq18}, \eq{eq24}-\eq{eq25}). At $\beta=\frac{2}{3}$ the critical exponent $\gamma(c)$ is plotted in \fig{fig06} against $c$ for large $c$.

The numerical evaluation of the variance in \eq{eq50b} depicted in \fig{fig06} clearly shows that for chosen $\beta=\frac{2}{3}$ and large $c$ (where $c=N/n$), the scaling exponent $\gamma(c)$ in \eq{eq50b} tends to the value $\gamma\approx 0.31$ which is very close to the KPZ exponent, $\nu=\frac{1}{3}$. To extrapolate the numerical results for $\gamma(c)$ for large $c$ we have used the Pad\'{e} approximant, fitting the points in \fig{fig07} by the ratio of two polynomials of the first order,
$$
\gamma(c)\approx \frac{A+Bc}{D + c}\bigg|_{c\to \infty} \to B
$$
The best fitting curve corresponds to the values $A=58.14, B=0.31, D=77.13$. For values of $\beta$ other than $\beta=\frac{2}{3}$ we do not see any good convergence for the critical exponent, which could mean that the scaling dependence $\left(\mathrm{Var}\,k(n)\right)^{1/2} = n^{\gamma}$ is violated, however this question demands further investigation.

Let us make a short comment concerning the possible critical transition for the ensembles of the torus knots. It is known that that the partition function for weighted magnetic Dyck paths in the $n\times m$ for prime ($n,m$) rectangular provides the invariants of the torus $T_{n,m}$ knots \cite{gorsky}. If $n=km$ for integer $k$ the invariants of the torus links are relevant instead. The critical behavior for ensemble of $T_{n,n+1}$ torus knots has been discussed in \cite{bgk}. Since our framework for critical behavior in conditional  magnetic Dyck paths is similar one could question if some analogue of transition from the Gaussian to TW fluctuations can be recognized at the torus knot or link side for some observable.

We make some preliminary remarks postponing more detailed analysis for the separate study. Since we consider the Dyck paths on the strip one "quantum number" of the torus knot is fixed by the width of the strip, say n. Recall that we are interested in arriving at height $k$ at $N$ steps Dyck path in the case when $N>n$. That is some number of bounces occurs at the path but we do not
add the corresponding weights. If $N\rightarrow \infty$ at fixed n we could say that we are looking at invariants of $T_{n,\infty}$ torus knots. However after all we consider $N=cn$  at $N\rightarrow \infty$ limit that is effectively we are looking at the so called stable limit of the torus knots $T_{\infty,\infty}$ but in the special limit when  $\frac{n}{m}={\rm const}$. The expected phase transition occurs at some critical $c=c^*$. Hence the meaning of parameter c can be identified. However it is not so simple to identify the analogue of the conditional probability we have used as indicator of the phase transition above. We could speculate that is corresponds to the counting of invariants of some embedding of simpler knots parametrized by $k$ into the stable $(\infty,\infty)$ limit. Certainly this issue deserves the additional study.

\section{Inflated random ring}

The classical problem in statistics of ideal polymers posed and solved in 1962 by M. Fixmann \cite{fixm} deals with the computation of the partition function, $Z_N(R_g)$, of $N$-step random walk with the gyration radius, $R_g$. Later, the same problem has been rediscovered independently by many researchers using variety of approaches \cite{fix-f1,fix-f2}. Here we consider the problem of calculating the distribution function $Z_N(r|R_g)$ of a particular monomer located at the point ${\bf r}$ in a Brownian ring with a fixed gyration radius, $R_g$. We pay attention to a very specific limit of inflated paths, when the gyration radius, $R_g$, scales with the chain length, $N$, super-diffusively, i.e. as $R_g = a N^{\alpha}$ ($\frac{1}{2}<\alpha\le 1$). In particular, we are interested in the limiting case, $R_g = a N$ of strongly inflated chain configurations.

The goal of our consideration is to emphasize the role of path stretching and of a particular role of imposed geometric constraints. We obtain the requested partition function in two different way: (i) by exact summation of all fluctuational modes of paths and (ii) by cutting-off only the "typical" fluctuational modes for stretched trajectories. We demonstrate the difference in the fluctuational behavior of trajectories: in the regime (i) the fluctuations are Gaussian, while in the regime (ii) they are controlled by the kPZ exponent $\nu=\frac{1}{3}$. At the end of Section we discuss the physical implementation of the model of inflated polymer ring, which permits to justify the cutting off the part of fluctuational modes.

\begin{figure}[ht]
\centering
\includegraphics[width=14cm]{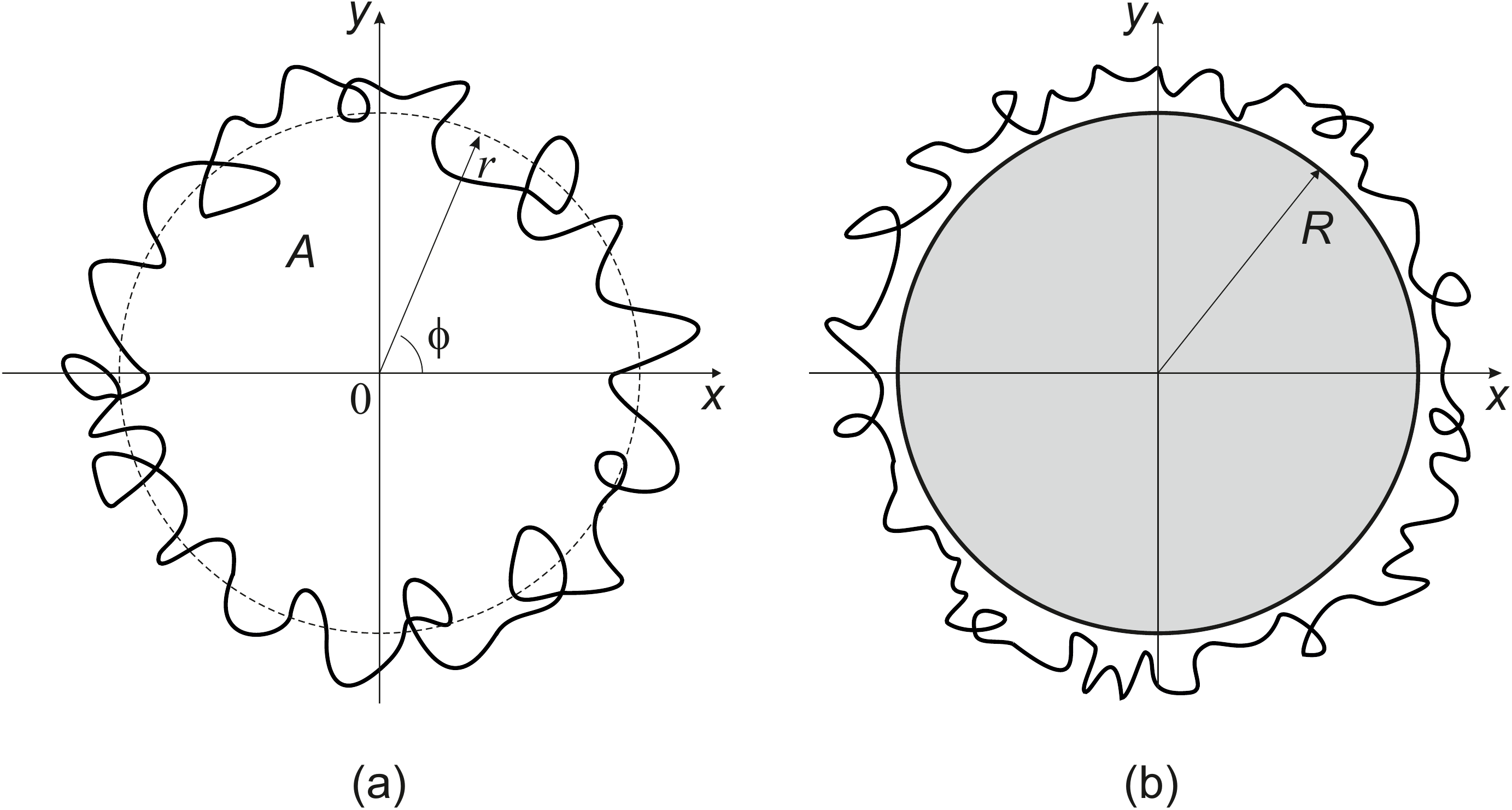}
\caption{(a) Inflated random ring with fixed gyration radius, $R_g$; (b) Inflated random ring is leaning on the impenetrable disc of radius, $R_g$.}
\label{fig07}
\end{figure}

The system under consideration is shown in \fig{fig07}a. Namely, we have the ideal ring polymer in the center-of-mass frame, where the center of mass,
$$
{\bf R}_c=\frac{1}{N}\sum_{j=1}^N {\bf r}_j
$$
is located at the origin of the 3D space, i.e. ${\bf R}_c=0$. By definition, the gyration radius, $R_g^2$, is:
\be
R_g^2=\frac{1}{2N^2}\sum_{j\neq k}^N ({\bf r}_j-{\bf r}_k)^2 = \frac{1}{N}\sum_{j=1}^N
\mathbf{r}_j^2-\left(\frac{1}{N}\sum_{j=1}^N\mathbf{r}_j\right)^2 = \frac{1}{N}\sum_{j=1}^N
\mathbf{r}_j^2-{\bf R}_c^2 \equiv \frac{1}{N}\sum_{j=1}^N \mathbf{r}_j^2
\label{eq51}
\ee
where in \eq{eq51} it is implied that ${\bf R}_c=0$.

Let us fix the "inflation degree" of the ideal ring chain by fixing the typical square of the gyration radius, $R_g^2$, in the grand canonical ensemble introducing the Lagrange multiplier, $s$. Define the grand canonical partition function:
\begin{multline}
Z_N(s) = \int d{\bf r}_1...d{\bf r}_{N-1} \exp\left(-u\sum_{j=1}^{N-1} ({\bf
r}_{j+1}-{\bf r}_j)^2\right) e^{- s R_g^2} \\ =
\int d{\bf r}_1...d{\bf r}_{N-1} \exp\left(-u\sum_{j=1}^{N-1} ({\bf
r}_{j+1}-{\bf r}_j)^2 -\frac{s}{N}\sum_{j\neq k}^N {\bf r}^2_j\right)
\label{eq52}
\end{multline}
where it is supposed that ${\bf r}_N={\bf r}_1$. Since all dimensions are independent, we can further proceed with one-dimensional system only.

The path integral formulation of \eq{eq52} reads
\be
Z_N(s) =\int {\cal D}\{x\}\exp\left\{-\int_0^N \left(u\, \dot{x}^2(t) + \frac{s}{N}\, x^2(t)\right)dt\right\} = \int {\cal D}\{x\} e^{-S}
\label{eq53}
\ee
where $\disp \dot{x} =\frac{\partial x(t)}{\partial t}$ and $0\le t \le N$. The Lagrangian $L$ of the action $S=\int_0^N L\, dt$ in \eq{eq53} is defined as
\be
L=u\,\dot{x}^2 + \frac{s}{N}\, x^2(t)
\label{eq54}
\ee
and the corresponding nonstationary Schr\"odinger-like equation for the probability distribution in parabolic well $V(x) = -\frac{s}{N}x^2$ is
\be
\frac{\partial P(x,t)}{\partial t} = \frac{1}{4u} \frac{\partial^2 P(x,t)}{\partial x^2} -  \frac{s}{N}\, x^2\, P(x,t)
\label{eq55}
\ee

In order to fix a "dictionary", it seems instructive to compare \eq{eq55} with the problem of quantum particle in a harmonic potential. The solution of a stationary quantum-mechanical problem
\be
- E_n\psi_n(x)=\frac{\hbar^2}{2m}\frac{\partial^2}{\partial x^2}\psi_n(x) - \frac{1}{2} m\omega^2 x^2 \psi_n(x)
\label{eq56}
\ee
reads
\be
\psi_n(x)=\left(\frac{m\omega}{\pi\hbar}\right)^{1/4}\frac{1}{\sqrt{2^n\, n!}}\,
H_n\left(x\sqrt{\frac{m\omega}{\hbar}}\right)\exp\left(-\frac{m\omega}{2\hbar}x^2\right); \qquad E_n=\left(n+\frac{1}{2}\right)\hbar \omega
\label{eq57}
\ee
where $H_n(...)$ is the Hermite polynomial.

Seeking now for the solution of \eq{eq55} we get
\be
\begin{cases}
\disp -\lambda_n W_n(x) = \frac{\partial W_n(x)}{\partial t} \medskip \\
\disp -\lambda_n W(x) = \frac{1}{4u} \frac{\partial^2 W_n(x)}{\partial x^2} - \frac{s}{N}\, x^2\, W_n(x)
\end{cases}
\label{eq58}
\ee
The correspondence between the parameters of \eq{eq55} and \eq{eq56} is set as follows
\be
E_n \leftrightarrow \lambda_n, \qquad
\frac{\hbar^2}{2m} \leftrightarrow \frac{1}{4u}, \qquad
\frac{1}{2} m\omega^2 \leftrightarrow \frac{s}{N}
\label{eq59}
\ee
The solution of the stationary polymer problem \eq{eq58} is
\be
W_n(x) = \left(\frac{su}{4N}\right)^{1/4}\frac{1}{\sqrt{2^n n!}} H_n\left(x\left(\frac{4su}{N}\right)^{1/4} \right)\exp\left(-x^2\left(\frac{su}{N}\right)^{1/2}\right); \quad \lambda_n=\left(n+\frac{1}{2}\right)\sqrt{\frac{s}{N}}
\label{eq60}
\ee

We are interested in the following distribution function
\be
Q(x,N) = \sum_{n=0}^{\infty}e^{-\lambda_n N}\, W_n^2(x)
\label{eq61}
\ee
where the eigenfunction $W_n(x)$ and corresponding eigenvalue $\lambda_n$ are defined in \eq{eq60}. The partition function $Q(x,N)$ given by \eq{eq60}--\eq{eq61} is the starting point for our further consideration.

\subsection{Exact summation of the series \eq{eq61}}

The distribution function $Q(x,N)$ in \eq{eq61} can be explicitly written as the following sum
\begin{multline}
Q(x,N)=\left(\frac{s u}{4N}\right)^{1/2}\exp\left(-2x^2\left(\frac{su}{N}\right)^{1/2}\right)\exp\left(-\frac{1}{2}\sqrt{s N}\right) \\ \times \sum\limits_{n=0}^\infty \frac{H^2_n\left(x \left(\frac{4 s u}{N}\right)^{1/4}\right)}{n!}\left(\frac{\exp\left(-\sqrt{s N}\right)}{2}\right)^n
\label{eq62}
\end{multline}
Now, using the properties of sums involving Hermite functions, we can write
\be
\sum\limits_{n=0}^\infty \frac{H_n\left(x\right)H_n\left(y\right)}{n!}\left(\frac{u}{2}\right)^n=\frac{1}{\sqrt{1-u^2}}
\exp\left(\frac{2u}{1+u}xy-\frac{u^2}{1-u^2}(x-y)^2\right); \qquad \vert u\vert <1
\label{eq63}
\ee
Performing the summation in \eq{eq63}, we obtain the expression for $Q(x,N)$:
\begin{multline}
Q(x,N)=\left(\frac{s u}{4N}\right)^{1/2}\exp\left(-2x^2\left(\frac{su}{N}\right)^{1/2}\right)\exp\left(-\frac{1}{2}\sqrt{s N}\right) \\ \times \frac{1}{\sqrt{1-\exp\left(-2\sqrt{s N}\right)}}\exp\left(\frac{4 \exp\left(-\sqrt{s N}\right)}{1+\exp\left(-\sqrt{s N}\right)}x^2 \left(\frac{s u}{N}\right)^{1/2}\right)
\label{eq64}
\end{multline}
It can be easily seen that the function $Q(x,N)$ possess Gaussian fluctuations.

\subsection{Saddle point estimation of the series \eq{eq61}}

Let us estimate the sum in \eq{eq61} via the saddle point method. To proceed, recall that $s$ is the Lagrange multiplier of $R_g^2$. Thus, to fix "softly" the trajectories with given $R_g^2$, we set
\be
s=R_g^{-2}
\label{eq65}
\ee
Later on we will "inflate" trajectory by increasing $R_g$ and simultaneous rescaling of $N$, which will be $N=a R_g^{\alpha}$, where $1\le \alpha <2$ and $a=\mathrm{const}$.

From \eq{eq61} we see that the dominant contribution to $Q(x,N)$ comes from such $n$, for which $\lambda_n\, N \approx 1$. Plugging the expression $\lambda_n = N^{-1}$ into \eq{eq60}, we arrive at the equations which determines the values of $n$ which give the main contribution to $Q(x,N)$:
\be
\frac{1}{N} = \left(n+\frac{1}{2}\right)\sqrt{\frac{s}{N}}
\label{eq66}
\ee
Solving \eq{eq66} at $n\gg 1$, and using \eq{eq65}, we get
\be
n=n^* \approx \frac{1}{\sqrt{sN}}=\frac{R_g}{\sqrt{N}}
\label{eq67}
\ee

Expressing all parameters in terms of $R_g$ and $N$, we can rewrite \eq{eq60} as follows
\be
W_n(x) = \left(\frac{u}{4R_g^2 N}\right)^{1/4}\frac{1}{\sqrt{2^n n!}} H_n\left(x\left(\frac{4u}{R_g^2 N}\right)^{1/4} \right)\exp\left(-x^2\left(\frac{u}{R_g^2 N}\right)^{1/2}\right)
\label{eq68}
\ee
It is known that the Hermite polynomial $H_n(z)$ at $z\approx 2\sqrt{n}$ and $n\gg 1$ have the asymptotic expansion
\be
H_n(z)\approx \sqrt{2\pi}\, 2^{-n/2} \exp\left(\frac{n\ln(2n)}{2}-\frac{3n}{2}+z\sqrt{n}\right)
n^{1/6}\mathrm{Ai}\left(\frac{z-2\sqrt{n}}{n^{-1/6}}\right); \quad z=x\left(\frac{4u}{R_g^2 N}\right)^{1/4}
\label{eq69}
\ee
The condition $z\approx 2\sqrt{n^*}$ sets the equation for $x=x^*$, at which the Airy tail of the Hermite polynomials appear
\be
x^{*}\left(\frac{4u}{R_g^2 N}\right)^{1/4} = 2 \left(\frac{R_g^2}{N}\right)^{1/4}
\label{eq70}
\ee
Corresponding expression with fluctuations, $\disp \frac{z-2\sqrt{n}}{n^{-1/6}}$, reads
\be
\frac{\disp x^{*}\left(\frac{4u}{R_g^2 N}\right)^{1/4} - 2 \left(\frac{R_g^2}{N}\right)^{1/4}} {\disp \left(\frac{R_g}{\sqrt{N}}\right)^{-1/6}} = \xi
\label{eq71}
\ee
where $\xi$ does not depend on $R_g$ and $N$. Solving \eq{eq71}, we get
\be
x^* = \left(\frac{2}{u}\right)^{1/4} R_g + \frac{\xi}{(4u)^{1/4}}\, (R_g\, N)^{1/3}
\label{eq72}
\ee

\subsection{What does the difference between \eq{eq64} and \eq{eq69} tell us?}

The full summation of the series \eq{eq61} ensures the Gaussian distribution (even for stretched paths), while the saddle-point approximation of \eq{eq61} for stretched trajectories provides the KPZ-like scaling for fluctuations. The cutoff in \eq{eq61} of modes with small eigenvalues $\lambda_k$ (i.e. of large wavelengths) does not permit the inflated ring possess large-scale fluctuations. From that point of view the cutoff looks like introducing the solid constraint in a form of an impenetrable disc as shown in \fig{fig07}. Similar behavior has been found in \cite{polov} for fluctuations of stretched random walk located in the vicinity of the impenetrable disc.

It seems that presence of the impenetrable disc restricting fluctuations of the Brownian ring is crucial for the localization of trajectories within the strip of width $R^{1/3}$. In both cases (the full summation and the saddle point approximation) the trajectories are pushed to an improbable tiny region of the phase space, however the presence of a large deviation regime seems not to be a sufficient condition to affect the statistics and the presence of the solid convex boundary on which trajectories is leaning, is crucial. The importance of a solid convex boundary has been investigated earlier in \cite{peres,shlosman} and has been recently poinred out in \cite{shlosman2}.

\section{Conclusion}

In the paper we have collected three different "mean-field-like" models involving Brownian excursions, which possess for fluctuations the Kardar-Parisi-Zhang scaling with the critical exponent $\nu=\frac{1}{3}$. The main message of our work is as follows. We have considered simple one-body systems in which imposed external constraints push the underlying stochastic process to a large deviation regime with the anomalous statistics. Specifically, we dealt with various incarnations of "stretched" trajectories in the non-uniform geometry.

In Section II we have considered the $N$-step Brownian excursions on finite super-trees of $K$ generations with linearly growing branching. It has been shown that implying the condition $N = c K$, the system experiences in the thermodynamic limit $N\to \infty$ the phase transition from Gaussian (at $c<c^*$) to KPZ-like (at $c>c^*$) fluctuations. The construction of the mean-field-like theory of that transition is our forthcoming goal.

We have considered the super-tree with the linear growing branching corresponding to the Hilbert space of harmonic oscillator. However the supertree representation of Hilbert space for a generic Hamiltonian system in Krylov basis is possible. In this case the Lancosz coefficients provide the tridiagonal representation of Liouvillian and therefore branching pattern on the super-tree. It was shown recently that  the dependence of the branching on the level of the tree is governed by the open Toda chain \cite{dg}. Moreover the particular solution to the classical Toda chain tells if a generic Hamiltonian system is chaotic at the quantum level or not. It would be interesting to discuss the possible transition from the Gaussian to TW statistics in more general system
using the findings for the generic tri-diagonal representations in the Krylov basis from \cite{dg}.

In Section III we have investigated the statistics of $N$-step (1+1)-dimensional Dyck excursions in a strip of finite width, $n$. We have forced Dyck paths to be inflated or compressed by assigning to each excursion the area below the curve. The area is controlled by the conjugated variable, $q$, in the grand canonical ensemble. As in the Section II, we have imposed the linear constraint $N=cn$, and have considered the variation of the middle point fluctuations in ensemble of Dyck paths at large $n$ and at fixed ratio $c=N/n$. We have found that for special selection of $q$, namely for $q=1+N^{-2/3}$ the fluctuations scale with the critical exponent $\gamma\approx 0.31$ at $c\gg 1$.

In Section IV we have studied the spatial fluctuations of bonds of the $N$-bond inflated polymer ring with fixed radius of gyration, $R_g$. The degree of the polymer inflation is controlled by the relation between $N$ and $R_g$. The problem goes back to early 60s of the last century and is a slight modification of the problem solved by M. Fixmann who was interested in the computation of the partition function, $Z(R_g,N)$ of the $N$-bond ideal polymer ring with fixed $R_g$. We have shown that cutting off the long-ranged spatial fluctuations and leaving only the "typical" modes in the stretched path with $R_g=aN$, we ensure the KPZ-like scaling for bond fluctuations. To the contrary, summing up all normal modes, we find the Gaussian fluctuations.

In all considered models the KPZ fluctuations emerge in presence of two complementary conditions: (i) the trajectories are pushed to a large deviation region of the phase space, and (ii) the trajectories are leaning on an impenetrable (nonconcave) boundary. It seems that both conditions are simultaneously responsible for KPZ-like fluctuations in one-body statistical systems.

\begin{acknowledgments}
We are grateful to N. Brilliantov, B. Meerson, G. Oshanin, K. Polovnikov, T. Prellberg, S. Shlosman, M. Tamm, and A. Vladimirov for numerous discussions and valuable comments. The work of was partially supported by the RFBR grant 18-29-13013. All authors acknowledge the BASIS foundation for the support in the frameworks of grants (17-11-122-1 for AG and 19-1-1-48-1 for SN and AV).
\end{acknowledgments}


\begin{thebibliography}{99}

\bibitem{kpz} M. Kardar, G. Parisi, and Y.-C. Zhang, Dynamic Scaling of Growing Interfaces, Phys. Rev. Lett. {\bf 56} 889 (1986)

\bibitem{halpin} T. Halpin-Healy and Y.-C. Zhang, Kinetic roughening phenomena, stochastic growth, directed polymers and all that, Physics Reports {\bf 254} 215 (1995)

\bibitem{johansson} K. Johansson, Discrete Polynuclear Growth and Determinantal Processes, Comm. Math. Phys. {\bf 242} 277 (2003)

\bibitem{spohn} M. Pr\"{a}hofer and H. Spohn, Universal Distributions for Growth Processes in 1+1 Dimensions and Random Matrices, Phys. Rev. Lett.  {\bf 84} 4882 (2000); M. Pr\"ahofer, H. Spohn, Scale Invariance of the PNG Droplet and the Airy Process, J. Stat. Phys. {\bf 108} 1071 (2002)

\bibitem{tw} C.A. Tracy and H. Widom, Level-spacing distributions and the Airy kernel, Commun. Math. Phys. {\bf 159} 151 (1994)

\bibitem{miettinen} L. Miettinen, M. Myllys, J. Merikoski, and J. Timonen, Experimental determination of KPZ height-fluctuation distributions, Eur. Phys., J. B {\bf 46}, 55 (2005)

\bibitem{takeuchi1} K. A. Takeuchi, M. Sano, Universal Fluctuations of Growing Interfaces: Evidence in Turbulent Liquid Crystals, Phys. Rev. Lett. {\bf 104}, 230601 (2010)

\bibitem{takeuchi2} K. A. Takeuchi, M. Sano, T. Sasamoto, and H. Spohn, Growing interfaces uncover universal fluctuations behind scale invariance, Sci. Rep. (Nature) {\bf 1}, 34 (2011)

\bibitem{majumdar} S.N. Majumdar and G. Schehr, Top eigenvalue of a random matrix: large deviations and third order phase transition	J. Stat. Mech. P01012 (2014)

\bibitem{baruch} N.R. Smith and B. Meerson, Geometrical optics of constrained Brownian excursion: from the KPZ scaling to dynamical phase transitions, J. Stat. Mech. (2019) 023205

\bibitem{baruch2} B. Meerson and N.R. Smith, Geometrical optics of constrained Brownian motion: three short stories, J. Phys. A: Math. Theor. {\bf 52}, 415001 (2019)

\bibitem{val} A.S. Gorsky, S.K. Nechaev, and A.F. Valov, On statistical models on supertrees, J. High Energ. Phys. {\bf 218}, 123 (2018)

\bibitem{edelman} I. Dumitriu and A. Edelman, Matrix models for beta ensembles, J. Math. Phys. {\bf 43} 5830 (2002)

\bibitem{17} A. Maritan, Random walk and the ideal chain problem on self-similar structures, Phys. Rev. Lett. {\bf 62} 2845 (1989)

\bibitem{ternovsky} F.F. Ternovsky, I.A. Nyrkova, and A.R. Khokhlov, Statistics of an ideal polymer chain near the bifurcation region of a narrow tube, Physica A {\bf 184} 342 (1992)

\bibitem{burda} Z. Burda, J. Duda, J.-M. Luck, and B. Waclaw, Localization of the maximal entropy random walk, Phys. Rev. Lett. {\bf 102} 160602 (2009)

\bibitem{heavy} S.K. Nechaev, M.V. Tamm, and O.V. Valba, Path counting on simple graphs: from escape to localization, J. Stat. Mech. 053301 (2017)

\bibitem{Kornyik} M. Kornyik and G. Michaletzky, Wigner matrices, the moments of roots of Hermite polynomials and the semicircle law, J. Approx. Theor. {\bf 211} 29 (2016)

\bibitem{krefl}  D. Kreﬂ, Non-perturbative quantum geometry II, JHEP {\bf 1412} 118 (2014); D. Kreﬂ, Non-Perturbative Quantum Geometry, JHEP {\bf 1608} 020 (2016)

\bibitem{Dominici} D. Dominici, Asymptotic analysis of the Hermite polynomials from their differential–difference equation, J. Diff. Eq. Appl. {\bf 13} 1115 (2007)

\bibitem{carlitz} L. Carlitz and J. Riordan, Two element lattice permutation numbers and their q
-generalization, Duke J. Math. {\bf 31} 371 (1964); J. F\"urlinger and J. Hofbauer, q-Catalan numbers, J. Comb. Theor. A {\bf 40} 248 (1985)

\bibitem{prellberg0} T. Prellberg and R. Brak, Critical exponents from nonlinear functional equations for partially directed cluster models, J. Stat. Phys. {\bf 78} 701 (1995)

\bibitem{rich1} C. Richard, A. J. Guttmann, and I. Jensen, Scaling function and universal amplitude combinations for self-avoiding polygons, J.Phys. A: Math. Gen. {\bf 34} L495 (2001)

\bibitem{rich2} C. Richard, Scaling behaviour of two-dimensional polygon models, J. Stat. Phys., {\bf 108} 459 (2002)

\bibitem{prellberg} A. Owczarek and T. Prellberg,  Enumeration of area-weighted Dyck paths with restricted height, Australasian Journal of Combinatorics {\bf 54}, 13, (2012)

\bibitem{quassem} Q.M. Al-Hassan, On powers of general tridiagonal matrices, Appl. Math. Sci., {\bf 9}, 583, (2015)

\bibitem{gorsky} E. Gorsky, in \emph{Zeta functions in algebra and geometry}, Contemporary Mathematics (2012), $q,t$-Catalan numbers and knot homology, 213

\bibitem{bgn} Bulycheva, K.; Gorsky, A.; Nechaev, S., Critical behavior in topological ensembles, Physical Review D, 92, id.105006 (2015)

\bibitem{fixm} M. Fixman, Radius of Gyration of Polymer Chains, J. Chem. Phys., {\bf 36}, 306 (1962)

\bibitem{fix-f1} B.E. Eichinger, An Approach to Distribution Functions for Gaussian Molecules, Macromolecules {\bf 10}, 671 (1977)

\bibitem{fix-f2} J. Rudnik and G. Gaspari, The Shapes of Random Walks, Science, {\bf 237}, 384 (1987)

\bibitem{polov} S. Nechaev, K. Polovnikov, S. Shlosman, A. Valov, and A. Vladimirov, Anomalous one-dimensional fluctuations of a simple two-dimensional random walk in a large-deviation regime, Phys. Rev. E {\bf 99}, 012110 (2019)

\bibitem{peres} A. Hammond and Y. Peres, Fluctuation of a planar Brownian loop capturing a large area, Trans. Am. Math. Soc., {\bf 360}, 6197 (2008)

\bibitem{shlosman} D. Ioffe, S. Shlosman, and Y. Velenik, An invariance principle to Ferrari--Spohn diffusions. Comm. Math. Phys., {\bf 336}, 905 (2015)

\bibitem{shlosman2} A. Vladimirov, S. Shlosman, and S. Nechaev, Brownian flights over a circle,  arXiv:2002.09965

\bibitem{dg} A. Dymarsky, A. Gorsky, Toda chain flow in Krylov space, e-Print: 1912.12227

\end{thebibliography}
\end{document}